\documentclass[11pt, a4paper]{article}
\usepackage{jheppub}

\title{String theories on warped AdS backgrounds 
and integrable deformations of spin chains}

\author{Takashi Kameyama,}
\author{Kentaroh Yoshida}

\affiliation{Department of Physics, Kyoto University, 
Kyoto 606-8502, Japan} 

\emailAdd{kame@gauge.scphys.kyoto-u.ac.jp}
\emailAdd{kyoshida@gauge.scphys.kyoto-u.ac.jp}

\abstract{
We study integrable deformations of AdS/CFT by focusing upon three kinds of warped AdS$_3$ geometries, 
1) space-like warped AdS$_3$, 2) time-like warped AdS$_3$ and 3) null warped AdS$_3$\,.  
These geometries are embedded into type IIB supergravity solutions and are regarded as consistent string backgrounds. 
By restricting the classical motion of strings on the warped AdS$_3\times$S$^1$ subspace, 
the Landau-Lifshitz sigma models are derived by taking the fast-moving limit. 
The first two warped AdS$_3$ spaces correspond to anisotropic deformations 
of the $sl(2)$ spin chain and the last one to Jordanian deformations. 
After taking the continuum limit of the deformed spin chains with coherent states, 
the resulting theories agree with the Landau-Lifshitz sigma models obtained from the string-theory side. 
}


\arxivnumber{1304.1286}

\begin{document}

\maketitle

\section{Introduction}

The AdS/CFT correspondence \cite{M,GKP,W} is still the focus of attention 
though
it has been studied exhaustively for more than the decade. 
It is now supported by an enormous amount of evidence.
The discovery of the integrable structure behind AdS/CFT is a truly tremendous achievement 
toward the complete proof. The results on this issue 
have been accumulated in the great review \cite{review}. 

\medskip 

The next step is to consider integrable deformations of AdS/CFT. 
The usual AdS/CFT belongs to the rational class of integrable systems. 
For example, the XXZ models, which are anisotropic deformations of the XXX model, 
shed light on the deeper integrable structure like quantum affine algebras, 
while the XXX model is nothing but a degenerate limit of the XXZ models. One may expect 
that a similar thing would happen in the case of AdS/CFT and hence it is of importance to consider 
integrable deformations of AdS/CFT. 

\medskip 

In the recent, some works have been done in this direction. 
There are mainly two approaches. The one is based on $q$-deformations of S-matrix of 
the integrable spin chain \cite{BK,BGM,HHM,dLRT,Arutyunov}. 
The other is to consider deformations of target spaces of string sigma models, 
namely, warped AdS spaces and/or deformed spheres. For earlier works and recent progress 
along this direction, 
for example, see \cite{Cherednik,FR,BFP} and 
\cite{Wen,KY,ORU,KOY,KYhybrid,KY-Sch,KMY-QAA,BR}, respectively.  

\medskip 

We follow here the latter approach and study integrable deformations of AdS/CFT 
by focusing upon three kinds of warped AdS$_3$ geometries, 
1) space-like warped AdS$_3$ \cite{DLP}, 2) time-like warped AdS$_3$ \cite{RS,DIsrael} 
and 3) null warped AdS$_3$ \cite{null}. The third one is also known as 
three-dimensional Schr\"odinger spacetimes \cite{Son,BM}. 
For a summary of warped AdS$_3$\,, see \cite{warped}.   
All of these geometries are embedded into supergravity solutions and are regarded 
as consistent string backgrounds\footnote{ 
For space-like and time-like warped AdS$_3$\,, see \cite{DLP,CDR,Anninos,Levi,OU}. 
For the null case, see \cite{CS,Bobev,SS}. }.
By restricting the classical motion of strings on the warped AdS$_3\times$S$^1$ subspace, 
the Landau-Lifshitz sigma models are derived by taking the fast-moving string limit \cite{Kruczenski,ST}. 

\medskip 

The first two warped AdS$_3$ geometries correspond to anisotropic deformations of the $sl(2)$ spin chain.    
The last one corresponds to Jordanian deformations, which mean null-like deformations.  
After taking the continuum limit of the deformed spin chains with coherent states, 
the resulting theories agree with the Landau-Lifshitz sigma models 
obtained from the string-theory side. 

\medskip 

This note is organized as follows. Section 2 is the setup of the string action 
and overviews three kinds of warped AdS$_3$\,. 
In section 3 we argue the correspondence between strings on space-like warped AdS$_3\times$S$^1$ 
and anisotropic deformations of the $sl(2)$ spin chain. By taking a fast moving limit of the relativistic string action, 
the Landau-Lifshitz sigma model action is derived. Then the continuum limit of the deformed spin chains is taken 
with coherent states. The both results show the agreement. 
In section 4 we consider the time-like warped AdS$_3$ case. The analysis is almost the same as in section 3. 
In section 5 we consider the null warped AdS$_3$ case, where one has to consider Jordanian deformations 
of the $sl(2)$ spin chain. The agreement of the Landau-Lifshitz sigma models 
is shown again. Section 6 is devoted to conclusion and discussion. 
The convention of $sl(2)$ generators is summarized in Appendix A. 
The basic property of $sl(2)$ coherent states is introduced in Appendix B. 
The continuum limit of the undeformed part of the $sl(2)$ spin chain is reviewed in Appendix C. 

\section{Setup}

We first introduce the string action and fix our notation. 
Then three kinds of warped AdS$_3$ are summarized to make our explanation clear. 

\subsection{String action}

The string action we consider here is given by 
\begin{eqnarray}
S = -\frac{1}{4\pi \alpha'}\int\!\!d\tau d\sigma\,\eta^{\mu\nu}
\left[\partial_{\mu}X^M \partial_{\nu}X^N G_{MN} \right]\,, \nonumber 
\end{eqnarray}
where the string world-sheet coordinates are $\sigma^{\mu}=(\tau,\sigma)$ with $\eta_{\mu\nu}=(-1,+1)$\,. 
The target space metric is given by $G_{MN}$ and $X^M(\tau,\sigma)$ are the embedding coordinates 
of the string world-sheet into the target space. 
The periodic boundary condition is imposed for the 
spatial direction of the string world-sheet, 
\begin{eqnarray}
X^{M}(\tau,\sigma + 2\pi) = X^{M}(\tau,\sigma)\,. \nonumber 
\end{eqnarray}
It is necessary to take account of the Virasoro constraints,  
\begin{eqnarray}
	G_{MN}\partial_{\tau}X^{M}\partial_{\sigma}X^{N} = 0\,,
	\label{virasoro1}
\end{eqnarray}
\begin{eqnarray}
	G_{MN}(\partial_{\tau}X^{M}\partial_{\tau}X^{N} + \partial_{\sigma}X^{M}\partial_{\sigma}X^{N}) = 0\,,
	\label{virasoro2}
\end{eqnarray}
which will be important in reducing the system 
to the Landau-Lifshitz sigma models.

\subsection{Warped AdS$_3$ geometries} 

There are three kinds of warped AdS$_3$ geometries, 1) space-like warped AdS$_3$\,, 2) time-like warped AdS$_3$ 
and 3) null warped AdS$_3$\,. The metrics of them are listed below. 

\subsubsection*{1) space-like warped AdS$_3$}

The metric of space-like warped AdS$_3$ is given by  
\begin{eqnarray}
ds^2 &=& \dfrac{R^{2}}{4} \left[- \cosh^{2}{\rho}dt^{2} + d\rho^{2} + (1 - C)(du + \sinh{\rho}dt)^{2} + d\varphi^2 \right]\,,
\label{metric1}
\end{eqnarray}
where $C$ is a constant, deformation parameter. The three variables $t,\rho,u$ describe 
the usual AdS$_3$ with the radius $R$ in the global coordinates when $C =0$\,. 

\medskip 

The metric (\ref{metric1}) can be rewritten as 
\begin{eqnarray}
ds^2 = \frac{R^2}{2} \left[ {\rm Tr}(JJ) -2C {\rm Tr}(T^1J){\rm Tr}(T^1J) \right]\,, 
\label{metric-g1}
\end{eqnarray}
in terms of the left-invariant one-form $J = g^{-1}dg$ with the group element 
\begin{eqnarray}
g = \textrm{e}^{- t T^{0}}\textrm{e}^{\rho T^{2}}\textrm{e}^{u T^{1}}\,. 
\label{2.5}
\end{eqnarray} 
For the definition of $T^a$'s, see Appendix A. 
Note that the expression (\ref{metric-g1}) gives rise to the definition of space-like warped AdS$_3$\,,  
independently of the coordinates.

\subsubsection*{2) time-like warped AdS$_3$} 

The metric of time-like warped AdS$_3$ is given by 
\begin{equation}
ds^2 = \dfrac{R^{2}}{4} \left[\cosh^{2}{\rho}du^{2} + d\rho^{2} - (1 + C)(dt + \sinh{\rho}du)^{2} + d\varphi^2 \right]\,.
\label{metric2}
\end{equation}
With the group element 
\begin{eqnarray}
g =  \textrm{e}^{u T^{1}}\textrm{e}^{\rho T^{2}}\textrm{e}^{- t T^{0}}\,,
\end{eqnarray}
the metric (\ref{metric2}) can be rewritten as 
\begin{equation}
	ds^2 = \frac{R^2}{2} \left[ {\rm Tr}(JJ) -2C {\rm Tr}(T^0J){\rm Tr}(T^0J) \right]\,.
	\label{metric-g2}
\end{equation}

\subsubsection*{3) null warped AdS$_3$} 

The metric of null warped AdS$_3$ is given by  
\begin{eqnarray}
ds^2 &=& \dfrac{R^{2}}{4} \left[- 2\textrm{e}^{- \rho}dUdV + d\rho^{2} - C \textrm{e}^{-2\rho} dV^2 \right]\,.
\label{metric3}
\end{eqnarray}
The metric (\ref{metric3}) is written as  deformations with the $T^{-}$ component, 
\begin{equation}
	ds^2 = \frac{R^2}{2} \left[ {\rm Tr}(JJ) -2C {\rm Tr}(T^-J){\rm Tr}(T^-J) \right]\,. 
	\label{sch}
\end{equation}
When the group element $g$ is parametrized as 
\begin{eqnarray}
g =  \textrm{e}^{V T^{+}}\textrm{e}^{\rho T^{2}}\textrm{e}^{U T^{-}}\,,
\end{eqnarray}
the metric (\ref{metric3}) is reproduced.

\section{Space-like warped AdS$_3$ and deformed spin chains}

\subsection{Fast-moving string limit}

We consider the fast-moving limit of strings propagating on space-like warped AdS$_3\times$S$^1$\,. 
We first have to change the coordinates of the metric (\ref{metric1})\,. 
The reason is that the warped AdS$_3$ is expressed as a $U(1)$ fibration over AdS$_2$ 
and it is not convenient to take the limit. 

\medskip 

Let us take the following parametrization of $g$\,, instead of (\ref{2.5}), 
\begin{equation}
	g = \textrm{e}^{- (\phi + t)T^0}\textrm{e}^{2\rho T^1}\textrm{e}^{(\phi - t) T^0}\,. 
	\label{a1}
\end{equation}
Then, with an S$^1$ circle described by $\varphi$\,, 
the metric of the warped AdS$_3\times$S$^1$ is given by 
\begin{equation}
	\begin{split}
	ds^{2} &= R^{2}\Bigl[ - \cosh^{2}{\rho}dt^{2} + d\rho^{2} + \sinh^{2}{\rho}d\phi^{2} + d\varphi^{2} \\
& \qquad - \dfrac{C}{4} [2\cos(\phi - t) d\rho - \sin(\phi - t)\sinh 2\rho (d\phi + dt)]^{2}\Bigr]\,. 
	\label{a13}
	\end{split}
\end{equation}
Note that the deformation part looks more complicated than the metric (\ref{metric1})\,, 
but the metric (\ref{a13}) is convenient to take the fast-moving limit, as we will see below. 

\medskip

Next we perform the coordinate transformation
\begin{equation}
	\phi = \tilde{\phi} + t, \qquad \varphi = \tilde{\varphi} + t, \qquad \rho = \dfrac{1}{2}\tilde{\rho}\,, 
	\label{a15}
\end{equation}
and take the gauge $ t = \kappa \tau$\,. Then we take the following limit,
\begin{equation}
	\kappa \rightarrow \infty  \quad \textrm{with} \quad \kappa^{2}C, \quad \kappa\partial_{\tau}\tilde{\phi},  
\quad \kappa\partial_{\tau}\tilde{\varphi}, \quad \kappa\partial_{\tau}\tilde{\rho}:~~ \textrm{fixed}\,.
	\label{a25}
\end{equation}
Note that the above limit (\ref{a25}) contains an additional condition on $\kappa^2 C$ in comparison to the usual one \cite{Kruczenski,ST,SL2}. 
This condition is necessary to ensure the finiteness of the resulting action as mentioned in \cite{Wen}. 

\medskip 

After all, the resulting action of the warped AdS$_{3}\times$S$^1$ is given by 
\begin{equation}
	\begin{split}
		S &= \dfrac{R^{2}}{4\pi\alpha'} \int\!\! d\tau d\sigma\,\biggl[- \kappa^{2}C\sin^{2}\tilde{\phi}\sinh^{2}{\tilde{\rho}} 
+ \kappa\bigl[(\cosh{\tilde{\rho}} - 1)\partial_{\tau}\tilde{\phi}  + 2\partial_{\tau}\tilde{\varphi}\bigr] \\&
\qquad \qquad \qquad \qquad
		- \dfrac{1}{4}\bigl[(\partial_{\sigma}\tilde{\rho})^{2} + 2(\cosh\tilde{\rho} - 1) (\partial_{\sigma}\tilde{\phi})^{2}\bigr] 
- (\partial_{\sigma}\tilde{\varphi})^{2}\biggr]\,.
	\label{a21}
	\end{split}
\end{equation}
The Virasoro constraints are also changed under the limit (\ref{a25}). 
To the leading order in $\kappa$, the first Virasoro constraint (\ref{virasoro1}) becomes 
\begin{equation}
	0 = \kappa[(\cosh{\tilde{\rho}} - 1)\partial_{\sigma}\tilde{\phi} + 2\partial_{\sigma}\tilde{\varphi}]\,.
	\label{a26}
\end{equation}
By eliminating $\partial_{\sigma}\tilde{\varphi}$ from (\ref{a21}) with (\ref{a26}),  
the leading-order action is given by 
\begin{equation}
	\begin{split}
		S &= \dfrac{R^{2}}{4\pi\alpha'} \int\!\! d\tau d\sigma\,\biggl[- \kappa^{2}C\sin^{2}\tilde{\phi}\sinh^{2}{\tilde{\rho}} 
+ \kappa\bigl[(\cosh{\tilde{\rho}} - 1)\partial_{\tau}\tilde{\phi} + 2\partial_{\tau}\tilde{\varphi}\bigr]\\
		& \qquad \qquad \qquad\qquad 
- \dfrac{1}{4}\bigl[(\partial_{\sigma}\tilde{\rho})^{2} 
+ \sinh^{2}{\tilde{\rho}}(\partial_{\sigma}\tilde{\phi})^{2}\bigr]\biggr]\,.
	\label{a28}
	\end{split}
\end{equation}
Note that the action (\ref{a28}) is written in the non-relativistic form in the sense of the string world-sheet. 
That is, it contains only the first order derivative on the world-sheet time $\tau$\,. 

\medskip 

For later purpose, it is helpful to rewrite the action (\ref{a28}) by introducing new parameters 
\begin{equation}
	L \equiv \dfrac{R^{2}\kappa}{2\pi\alpha'}\,, \qquad \lambda \equiv \dfrac{R^{4}}{\alpha '^{2}}\,.
	\label{a29}
\end{equation}
Then the action is rewritten as\footnote{The similar action is derived also in \cite{Wen}. 
However, the deformation term we obtained is different from the one in \cite{Wen} by $\sin^2\tilde{\phi}$\,. 
This factor is necessary to see the agreement with the result from the continuum limit of the deformed $sl(2)$ spin chains.} 
\begin{equation}
	\begin{split}
	S &= \dfrac{L}{2} \int\!\! dt d\sigma\,\biggl[- C\sin^{2}\tilde{\phi}\sinh^{2}{\tilde{\rho}} 
+ \bigl[(\cosh{\tilde{\rho}} - 1)\partial_{t}\tilde{\phi} + 2\partial_{t}\tilde{\varphi}\bigr] \\
& \qquad\qquad\qquad\qquad -\dfrac{\lambda}{16\pi^{2}L^{2}}\bigl[(\partial_{\sigma}\tilde{\rho})^{2} 
+ \sinh^{2}{\tilde{\rho}}(\partial_{\sigma}\tilde{\phi})^{2}\bigr]\biggr]\,,
	\label{a30}
	\end{split}
\end{equation}
where the time variable has been changed from $\tau$ to $t$ through $t=\kappa\tau$\,. 
As a result, a new potential term has been added. 
When $C=0$\,, the result in \cite{ST,SL2} is reproduced.  

\medskip 

Finally let us see the equations of motion, 
\begin{align}
	C\sin\tilde{2\phi}\sinh^{2}\tilde{\rho} + \sinh{\tilde{\rho}}\partial_{t}\tilde{\rho} 
- \dfrac{\lambda}{8\pi^{2}L^{2}}\partial_{\sigma}(\partial_{\sigma}\tilde{\phi}\sinh^{2}\tilde{\rho}) &= 0\,,\\
	- C \sin^{2}\tilde{\phi}\sinh\tilde{2\rho} + \sinh\tilde{\rho}\partial_{t}\tilde{\phi} 
+ \dfrac{\lambda}{8\pi^{2}L^{2}}[\partial^{2}_{\sigma}\tilde{\rho} 
- \dfrac{1}{2}\sinh2\tilde{\rho}(\partial_{\sigma}\tilde{\phi})^{2}] &= 0\,.
	\label{a31}
\end{align}
These are identical to the Landau-Lifshitz equations\footnote{Note that 
	$(\vec{n} \times \vec{m})_{i} = \varepsilon_{ijk}n^{j}m^{k}\,.$ Here we have introduced the totally antisymmetric tensor $\varepsilon_{ijk}$ with $\varepsilon_{012} = + 1$ and the vector $\vec{n}$ parameterizes $sl(2)$ coherent states. The indices of the vector are raised and  lowered with $\eta^{ij} = \textrm{diag}(+1\,, -1\,, -1)\,.$ For the detail of $sl(2)$ coherent states, see Appendix B. }
\begin{eqnarray}
	-\, \partial_{t} \vec{n} =  \dfrac{\lambda}{8\pi^{2}L^{2}}\vec{n} \times \partial_{\sigma}^{2} \vec{n} + \vec{n} \times \mathcal{J} \vec{n}\,,
\end{eqnarray}
with the anisotropic matrix $\mathcal{J}$ 
\begin{eqnarray}
\mathcal{J}^{ij} =  \textrm{diag}(j, - (j - 2C), - j) \qquad (j :~\textrm{an~arbitrary~const.})\,.
\end{eqnarray}
Note that the new potential term in (\ref{a30}) is related to the following quantity,  
\begin{equation}
	\mathcal{J}^{ij} n_{i}n_{j} = j + 2C \sin^{2}\tilde{\phi}\sinh^{2}\tilde{\rho}\,.
\end{equation}

\subsection{From anisotropically deformed $sl(2)$ spin chains}

It is a turn to examine the continuum limit of the deformed $sl(2)$ spin chains with $sl(2)$ coherent states.

\medskip 

The Hamiltonian of anisotropically deformed $sl(2)$ spin chains is described as
\begin{equation}
	H = \dfrac{\lambda}{8\pi^{2}}\sum_{k = 1}^{L} H_{k\,k + 1} + \xi\,\sum_{k=1}^{L}S_{1, k} \otimes S_{1, k + 1}\,.
\label{3.18}
\end{equation}
For the undeformed part, $H_{k\,k + 1}$ describes the nearest-neighbor interactions in the $sl(2)$ spin chain and 
the prefactor $\lambda/8\pi^2$ is now chosen for later convenience. 
For the deformation part, the $\xi$ is a constant parameter and  
the variable $S_{1, k}$ denotes the $sl(2)$ generator in the infinite-dimensional representation 
at the $k$-th site. For the detail of $S_{i}$, see Appendix A. 

\medskip 

With the coherent state for the whole spin chain 
\begin{equation}
|\vec{n}\rangle \equiv \prod_{k = 1}^{L} |\vec{n}_{k}\rangle\,, 
\label{whole}
\end{equation}
the expectation value of $H$ is evaluated as 
\begin{eqnarray}
	 \langle\vec{n}|H|\vec{n}\rangle &=&  \dfrac{\lambda}{8\pi^{2}}\sum_{k = 1}^{L} \langle\vec{n}_{k}\vec{n}_{k + 1}|H_{k\,k + 1}|\vec{n}_{k}\vec{n}_{k + 1}\rangle 
	 \nonumber \\ 
&& + \xi\,\sum_{k=1}^{L} \langle\vec{n}_{k}|S_{1, k}|\vec{n}_{k}\rangle \langle\vec{n}_{k + 1}|S_{1, k + 1}|\vec{n}_{k + 1}\rangle \,.
\end{eqnarray}
The undeformed part is computed in Appendix C. The deformation part becomes 
\begin{equation}
	\begin{split}
	\langle\vec{n}_{k}|S_{1, k}|\vec{n}_{k}\rangle = \dfrac{1}{2}\sin\phi_{k}\sinh\rho_{k}\,.
	\label{d9}
	\end{split}
\end{equation}
For the derivation, see Appendix B. 

\medskip 

By taking the continuum limit, the expectation value of the deformed Hamiltonian with the coherent states 
is evaluated as 
\begin{equation}
	\langle\vec{n}|H|\vec{n}\rangle \quad \longrightarrow \quad 
L\int\!\! d\sigma\, \biggl[\dfrac{\lambda}{32\pi^{2}L^{2}}\bigl[(\partial_{\sigma}\rho)^{2} 
+ \sinh^{2}\rho(\partial_{\sigma}\phi)^{2}\bigr] + \dfrac{\xi}{4} \sin^{2}\phi\sinh^{2}\rho\biggr]\,.
	\label{d10}
\end{equation}
The deformation part contributes to the Hamiltonian as a non-derivative term. 

\medskip 

Then the Wess-Zumino term should be added to ensure the quantization condition at each site of the spin chain. 
Thus the total action is given by 
\begin{equation}
	\begin{split}
	S &= \dfrac{L}{2}\int\!\! dt d\sigma\, \biggl[- C\sin^{2}\phi\sinh^{2}\rho + (\cosh\rho - 1)\partial_{t}\phi \\
& \qquad\qquad\qquad\qquad 
- \dfrac{\lambda}{16\pi^{2}L^{2}}\bigl[(\partial_{\sigma}\rho)^{2} + \sinh^{2}\rho(\partial_{\sigma}\phi)^{2}\bigr]\biggr]\,,
	\label{d11}
	\end{split}
\end{equation}
through the identification $\xi = 2C$\,. The action (\ref{d11}) agrees with the Landau-Lifshitz sigma models 
obtained from the string-theory side, up to the S$^1$ circle. 

\section{Time-like warped AdS$_3$ and deformed spin chains}

\subsection{Fast-moving string limit}

We consider the string action on time-like warped AdS$_3\times$S$^1$ here. 
As in the case of space-like warped AdS$_3$\,, we first have to change the coordinate system of the metric (\ref{metric2}). 
Again, we start from the metric (\ref{metric-g2}) written with the group element and take the parametrization (\ref{a1})\,.  
Then the metric of the warped AdS$_3\times$S$^1$ is given by
\begin{eqnarray}
	ds^{2} = R^{2}\Bigl[ - \cosh^{2}{\rho}dt^{2} + d\rho^{2}+ \sinh^{2}{\rho}d\phi^{2} + d\varphi^{2}
- C \left(\cosh^{2}\rho\, dt + \sinh^{2}\rho\, d\phi\right)^{2}\Bigr]\,,
	\label{f2}
\end{eqnarray}  
where an S$^1$ circle is described by $\varphi$\,.

\medskip 

We perform here the same coordinate transformation as in (\ref{a15}) with the gauge	$t = \kappa \tau$\,.
Then the fast-moving limit (\ref{a25}) is taken for the same reason as mentioned in the previous.
The resulting action of the warped AdS$_{3}\times$S$^1$ is given by 
\begin{equation}
	\begin{split}
		&S = \dfrac{R^{2}}{4\pi\alpha'} \int\!\! d\tau d\sigma\,
\biggl[- \kappa^{2}C\cosh^{2}\tilde{\rho} + \kappa\bigl[(\cosh{\tilde{\rho}} - 1)
\partial_{\tau}\tilde{\phi} + 2\partial_{\tau}\tilde{\varphi}\bigr]\\
		& \qquad \qquad \qquad \quad \quad
- \dfrac{1}{4}\bigl[(\partial_{\sigma}\tilde{\rho})^{2} +  2(\cosh{\tilde{\rho}} - 1)
(\partial_{\sigma}\tilde{\phi})^{2}\bigr] - (\partial_{\sigma}\tilde{\phi})^{2} \biggr]\,.
	\label{timelikeaction}
	\end{split}
\end{equation}

\medskip 

The next task is to examine the Virasoro constraints under the limit (\ref{a25}).
To the leading order in $\kappa$, the first 
Virasoro constraint (\ref{virasoro1}) becomes
\begin{equation}
	0 = \kappa[(\cosh{\tilde{\rho}} - 1)\partial_{\sigma}\tilde{\phi} + 2\partial_{\sigma}\tilde{\varphi}]\,.
	\label{f13}
\end{equation}
By eliminating $\partial_{\sigma}\tilde{\varphi}$ from (\ref{timelikeaction}) with (\ref{f13}),
the leading-order action is given by
\begin{equation}
	\begin{split}
		S &= \dfrac{R^{2}}{4\pi\alpha'} \int\!\! d\tau d\sigma\,\biggl[- \kappa^{2}C\cosh^{2}{\tilde{\rho}} + \kappa\bigl[(\cosh{\tilde{\rho}} - 1)\partial_{\tau}\tilde{\phi} + 2\partial_{\tau}\tilde{\varphi}\bigr] \\
		& \qquad \qquad \qquad \qquad
- \dfrac{1}{4}\bigl[(\partial_{\sigma}\tilde{\rho})^{2} + \sinh^{2}{\tilde{\rho}}(\partial_{\sigma}\tilde{\phi})^{2}\bigr]\biggr]\,.
	\label{f15}
	\end{split}
\end{equation}
Note that the action (\ref{f15}) is written in the non-relativistic form again. 

\medskip 

By introducing the following quantities,
\begin{equation}
	L \equiv \dfrac{R^{2}\kappa}{2\pi\alpha'}, \qquad \lambda \equiv \dfrac{R^{4}}{\alpha '^{2}}\,,
	\label{f16}
\end{equation}
the action is rewritten as
\begin{equation}
	\begin{split}
	S &= \dfrac{L}{2} \int\!\! dt d\sigma\,\biggl[- C\cosh^{2}{\tilde{\rho}} + \bigl[(\cosh{\tilde{\rho}} - 1)\partial_{t}\tilde{\phi} + 2\partial_{t}\tilde{\varphi}\bigr] \\
		& \qquad \qquad \qquad \qquad
		- \dfrac{\lambda}{16\pi^{2}L^{2}}\bigl[(\partial_{\sigma}\tilde{\rho})^{2} + \sinh^{2}{\tilde{\rho}}(\partial_{\sigma}\tilde{\phi})^{2}\bigr]\biggr]\,.
	\label{f17}
	\end{split}
\end{equation}
where the time variable has been changed from $\tau$ to $t$ through $t=\kappa\tau$\,. As a result, a new potential term 
has been added. 
When $C=0$\,, the result in \cite{ST,SL2} is reproduced again.  

\medskip 

Finally let us see the equations of motion,
\begin{align}
	\sinh{\tilde{\rho}}\partial_{t}\tilde{\rho} - \dfrac{\lambda}{8\pi^{2}L^{2}}\partial_{\sigma}(\partial_{\sigma}\tilde{\phi}\sinh^{2}\tilde{\rho}) &= 0\,,\\
	- C \sinh\tilde{2\rho} + \sinh\tilde{\rho}\partial_{t}\tilde{\phi} + \dfrac{\lambda}{8\pi^{2}L^{2}}[\partial^{2}_{\sigma}\tilde{\rho} - \dfrac{1}{2}\sinh2\tilde{\rho}(\partial_{\sigma}\tilde{\phi})^{2}] &= 0\,.
	\label{f18}
\end{align}
They are identical to the Landau-Lifshitz equations 
\begin{eqnarray}
	-\, \partial_{t} \vec{n} =  \dfrac{\lambda}{8\pi^{2}L^{2}}\vec{n} \times \partial_{\sigma}^{2} \vec{n} + \, \vec{n} \times \mathcal{J} \vec{n}\,,
\end{eqnarray}
with the anisotropic matrix $\mathcal{J}$
\begin{eqnarray}
\mathcal{J}^{ij} =  \textrm{diag}(j + 2C, - j,  -j) \qquad (j :~\textrm{an~arbitrary~const.})\,.
\end{eqnarray}
Note that the new potential term in (\ref{f17}) is related to the following quantity,  
\begin{eqnarray}
\mathcal{J}^{ij} n_{i}n_{j}=  j + 2C\cosh^{2}{\tilde{\rho}}\,.
\end{eqnarray}

\subsection{From anisotropically deformed $sl(2)$ spin chains}

Let us examine the continuum limit of the deformed $sl(2)$ spin chains with coherent states.

\medskip 

The Hamiltonian of anisotropically deformed $sl(2)$ spin chains is described as
\begin{equation}
	H = \dfrac{\lambda}{8\pi^{2}}\sum_{k = 1}^{L} H_{k\,k + 1} + \xi\,\sum_{k=1}^{L}S_{0, k} \otimes S_{0, k + 1}\,.
	\label{g1}
\end{equation}
This is similar to the Hamiltonian (\ref{3.18}). The only difference is that $S_{1,k}$ is replaced by $S_{0,k}$\,. 

\medskip 

With the coherent state (\ref{whole})\,,  
the expectation value of $H$ is evaluated as 
\begin{eqnarray}
	 \langle\vec{n}|H|\vec{n}\rangle &=&  \dfrac{\lambda}{8\pi^{2}}\sum_{k = 1}^{L} \langle\vec{n}_{k}\vec{n}_{k + 1}|H_{k\,k + 1}|\vec{n}_{k}\vec{n}_{k + 1}\rangle 
	 \nonumber \\ 
&& + \xi\,\sum_{k=1}^{L} \langle\vec{n}_{k}|S_{0, k}|\vec{n}_{k}\rangle \langle\vec{n}_{k + 1}|S_{0, k + 1}|\vec{n}_{k + 1}\rangle\,.
\end{eqnarray}
For the undeformed part, see Appendix C. 
The deformation part is computed like
\begin{equation}
	\begin{split}
	\langle\vec{n_{k}}|S_{0, k}|\vec{n_{k}}\rangle = \dfrac{1}{2}\cosh\rho_{k}\,, 
	\end{split}
\end{equation}
as shown in Appendix B. 

\medskip 

By taking the continuum limit, the expectation value 
is evaluated as 
\begin{equation}
		\langle\vec{n}|H|\vec{n}\rangle \quad \longrightarrow \quad 
L\int\!\! d\sigma\, \biggl[\dfrac{\lambda}{32\pi^{2}L^{2}}\bigl[(\partial_{\sigma}\rho)^{2} + \sinh^{2}\rho(\partial_{\sigma}\phi)^{2}\bigr] + \dfrac{\xi}{4} \cosh^{2}\rho\biggr]\,.
	\label{g4}
\end{equation}
The deformation part contributes to the Hamiltonian as a non-derivative term. 

\medskip 

Then the Wess-Zumino term should be added to ensure the quantization condition at each site of the spin chain. 
Thus the total action is given by 
\begin{equation}
	\begin{split}
	S &= \dfrac{L}{2}\int\!\! dt d\sigma\, \biggl[- C \cosh^{2}\rho + (\cosh\rho - 1)\partial_{t}\phi \\
& \qquad \qquad \qquad \qquad
- \dfrac{\lambda}{16\pi^{2}L^{2}}\bigl[(\partial_{\sigma}\rho)^{2} + \sinh^{2}\rho(\partial_{\sigma}\phi)^{2}\bigr]\biggr]\,,
	\label{g5}
	\end{split}
\end{equation}
through the identification $\xi = 2 C$\,.
The action (\ref{g5}) agrees with the Landau-Lifshitz sigma models obtained in the string-theory side, 
up to the S$^1$ circle. 

\section{Null warped AdS$_3$ and Jordanian deformations}

Finally we consider the case of strings propagating on null warped AdS$_3\times$S$^1$ here.  
We derive the Landau-Lifshitz sigma models from the string sigma models by taking the fast-moving limit.  
The resulting deformation term can also be correctly reproduced from Jordanian deformations 
of the $sl(2)$ spin chain by taking a continuum limit with coherent states.  

\subsection{Fast-moving string limit} 

For the case of null warped AdS$_3$\,, there is a technical difficulty. 
It is because that the metric (\ref{metric3}) is written with the Poincar$\acute{\textrm{e}}$ coordinates 
and we have to rewrite the metric (\ref{sch}) in terms of the global coordinate in the first place. 
The global coordinates for null warped AdS$_3$ is discussed in \cite{Blau}, 
but it does not make sense to consider the fast-moving string limit. 

\medskip 

A possible resolution is to start from the metric (\ref{sch}) with the group element 
and take the parametrization (\ref{a1})\,. 
Then  the metric of null warped AdS$_3\times$S$^1$ is given by 
\begin{align}
	ds^{2} &= R^{2}\Bigl[d\rho^{2} - \cosh^{2}{\rho}dt^{2} + \sinh^{2}{\rho}d\phi^{2} + d\varphi^{2} \notag \\
& \qquad \qquad 
-\dfrac{C}{2}\bigl[\cosh\rho(\cosh\rho + \sin(\phi - t)\sinh\rho)dt \notag \\
& \qquad \qquad \qquad \qquad
+ \sinh\rho(\sinh\rho + \sin(\phi - t)\cosh\rho)d\phi - \cos(\phi - t) d\rho\bigr]^{2}\Bigr]\,, 
	\label{b15}
\end{align} 
where an S$^{1}$ circle is parametrized by $\varphi$\,. 
Note that there is no difficulty to consider the fast moving string limit. 

\medskip 

Next we perform the same coordinate transformation as in (\ref{a15}) with the gauge $t = \kappa \tau$\,.
Then the fast-moving limit (\ref{a25}) is taken for the same reason as mentioned in the previous. 
The resulting action on the warped AdS$_3\times$S$^1$  is given by
\begin{align}
		&S = \dfrac{R^{2}}{4\pi\alpha'} \int\!\! d\tau d\sigma\, \biggl[- \kappa^{2}\dfrac{C}{2}(\cosh\tilde{\rho} + \sin\tilde{\phi}\sinh\tilde{\rho})^{2} + \kappa \bigl[(\cosh\tilde{\rho} - 1)\partial_{\tau}\tilde{\phi} + 2\partial_{\tau}\tilde{\varphi}\bigr] \notag \\
&\qquad \qquad \qquad \qquad 
- \dfrac{1}{4}\bigl[(\partial_{\sigma}\tilde{\rho})^{2}
+  2(\cosh\tilde{\rho} - 1) (\partial_{\sigma}\tilde{\phi})^{2} \bigr]
- (\partial_{\sigma}\tilde{\varphi})^{2}\biggr]\,. 
	\label{b23}
\end{align}
The Virasoro constraints  are also changed under the limit (\ref{a25}).
To the leading order in $\kappa$, the first Virasoro constraint (\ref{virasoro1}) becomes
\begin{equation}
	0 = \kappa[(\cosh{\tilde{\rho}} - 1)\partial_{\sigma}\tilde{\phi} + 2\partial_{\sigma}\tilde{\varphi}]\,.
	\label{b28}
\end{equation}
which can be used to solve $\partial_{\sigma}\tilde{\varphi}$ in terms of $\partial_{\sigma}\tilde{\phi}$.
After eliminating $\partial_{\sigma}\tilde{\varphi}$\,, 
to the leading-order action is given by
\begin{equation}
	\begin{split}
		S &= \dfrac{R^{2}}{4\pi\alpha'} \int\!\! d\tau d\sigma\,
\biggl[- \dfrac{\kappa^{2}C}{2}(\cosh\tilde{\rho} + \sin\tilde{\phi}\sinh\tilde{\rho})^{2} 
+ \kappa\bigl[(\cosh{\tilde{\rho}} - 1)\partial_{\tau}\tilde{\phi} + 2\partial_{\tau}\tilde{\varphi}\bigr]\\
& \qquad \qquad \qquad \qquad \qquad
		- \dfrac{1}{4}\bigl[(\partial_{\sigma}\tilde{\rho})^{2} + \sinh^{2}{\tilde{\rho}}(\partial_{\sigma}\tilde{\phi})^{2}\bigr]\biggr]\,.
	\label{b30}
	\end{split}
\end{equation}
Note that the action (\ref{b30}) is written in the non-relativistic form again. 

\medskip 

By introducing new quantities like 
\begin{equation}
	L \equiv \dfrac{R^{2}\kappa}{2\pi\alpha'}, \qquad \lambda \equiv \dfrac{R^{4}}{\alpha '^{2}}\,, 
	\label{b31}
\end{equation}
the following action is obtained, 
\begin{equation}
	\begin{split}
	S &= \dfrac{L}{2} \int\!\! dt d\sigma\,\biggl[ - \dfrac{C}{2}(\cosh\tilde{\rho} + \sin\tilde{\phi}\sinh\tilde{\rho})^{2} 
+ \bigl[(\cosh{\tilde{\rho}} - 1)\partial_{t}\tilde{\phi} + 2\partial_{t}\tilde{\varphi}\bigr]\\
& \qquad \qquad \qquad \qquad \qquad
- \dfrac{\lambda}{16\pi^{2}L^{2}}\bigl[(\partial_{\sigma}\tilde{\rho})^{2} 
+ \sinh^{2}{\tilde{\rho}}(\partial_{\sigma}\tilde{\phi})^{2}\bigr]\biggr]\,.
	\label{b32}
	\end{split}
\end{equation}
where the time variable has been changed from $\tau$ to $t$ through $t=\kappa\tau$\,. As a result, a new potential term has been added. 
When $C=0$\,, the result in \cite{ST,SL2} is reproduced.  

\medskip 

Finally let us see the equations of motion, which are derived from (\ref{b32}), 
\begin{equation}
	\begin{split}
	C\cos\tilde{\phi}\sinh\tilde{\rho}\,(\cosh\tilde{\rho} + \sin\tilde{\phi}\sinh\tilde{\rho}) 
+ \sinh{\tilde{\rho}}\,\partial_{t}\tilde{\rho} 
- \dfrac{\lambda}{8\pi^{2}L^{2}}\partial_{\sigma}(\partial_{\sigma}\tilde{\phi}\sinh^{2}\tilde{\rho}) &= 0\,,\\
	- C \Bigl[\sin\tilde{\phi} \cosh2\tilde{\rho} 
+ \dfrac{1}{2}(1 + \sin^{2}\tilde{\phi})\sinh2\tilde{\rho}\Bigr] 
+ \sinh\tilde{\rho}\,\partial_{t}\tilde{\phi} \hspace*{3.2cm} & \\
+ \dfrac{\lambda}{8\pi^{2}L^{2}} 
\Bigl[\partial^{2}_{\sigma}\tilde{\rho} 
- \dfrac{1}{2}\sinh2\tilde{\rho}(\partial_{\sigma}\tilde{\phi})^{2} \Bigr] &= 0\,. \notag
	\end{split}
\end{equation}
These are identical to the Landau-Lifshitz equations 
\begin{eqnarray}
	-\, \partial_{t} \vec{n} =  \dfrac{\lambda}{8\pi^{2}L^{2}}\vec{n} \times \partial_{\sigma}^{2} \vec{n} + \vec{n} \times \mathcal{J} \vec{n}\,,
\end{eqnarray}
with the anisotropic matrix $\mathcal{J}$
\begin{eqnarray}
\mathcal{J}^{ab} =  \begin{pmatrix}
			j + C & C & 0 \\
			C & - (j - C) & 0 \\
			0 & 0 & - j
			\end{pmatrix}
\qquad (j~: \textrm{an arbitrary const.})\,. 
\end{eqnarray}
Note that the new potential term in (\ref{b32}) is related to the following quantity,
\begin{eqnarray}
\mathcal{J}^{ij} n_{i}n_{j}=  j + C(\cosh\tilde{\rho} + \sin\tilde{\phi}\sinh\tilde{\rho})^{2} \,.
\end{eqnarray}

\subsection{From Jordanian deformations of the $sl(2)$ spin chain}

Let us take the continuum limit of  Jordanian deformed $sl(2)$ spin chains with coherent states.
The Jordanian deformed Hamiltonian is given by\footnote{Jordanian deformations of the XXX model are discussed 
in \cite{Jordanian}. The Hamiltonian (\ref{e1}) is the $sl(2)$ analogue of the deformed Hamiltonian.} 
\begin{equation}
	H = \dfrac{\lambda}{8\pi^{2}}\sum_{k = 1}^{L} H_{k\,k + 1} 
+ \dfrac{\xi}{2}\sum_{k = 1}^{L}\,(S_{0, k} + S_{1, k}) \otimes (S_{0, k + 1} + S_{1, k + 1})\,.
	\label{e1}
\end{equation}
This is similar to the Hamiltonian (\ref{3.18}) and (\ref{g1}). The only difference is that the variables in deformation part are replaced by $S_{0, k} + S_{1, k}$\,. 

\medskip 

Then the expectation value of $H$ 
with the coherent state (\ref{whole})
is evaluated as 
\begin{equation}
	\begin{split}
	 \langle\vec{n}|H|\vec{n}\rangle &=  \dfrac{\lambda}{8\pi^{2}}\sum_{k = 1}^{L} \langle\vec{n}_{k}\vec{n}_{k + 1}|H_{k\,k + 1}|\vec{n}_{k}\vec{n}_{k + 1}\rangle \\
	& \quad + \dfrac{\xi}{2}\,\sum_{k=1}^{L} \langle\vec{n}_{k}|(S_{0, k} + S_{1, k})|\vec{n}_{k}\rangle \langle\vec{n}_{k + 1}|(S_{0, k + 1} + S_{1, k + 1})|\vec{n}_{k + 1}\rangle\,. 
	\end{split}
\end{equation}
For the undeformed part, see Appendix C. The deformation part can be rewritten as 
\begin{equation}
	\begin{split}
	\langle\vec{n}_{k}| (S_{0, k} + S_{1, k})|\vec{n}_{k}\rangle = \dfrac{1}{2}(\cosh\rho_{k} + \sin\phi_{k}\sinh\rho_{k})\,,
	\nonumber 
	\end{split}
\end{equation}
with the help of the formula in Appendix B.

\medskip 

By taking the continuum limit, the expectation value 
is given by
\begin{equation}
	\langle\vec{n}|H|\vec{n}\rangle  \quad \longrightarrow \quad 
L\int\!\! d\sigma\, \biggl[\dfrac{\lambda}{32\pi^{2}L^{2}}\bigl[(\partial_{\sigma}\rho)^{2} 
+ \sinh^{2}\rho(\partial_{\sigma}\phi)^{2}\bigr] + \dfrac{\xi }{8} (\cosh\rho + \sin\phi\sinh\rho)^{2}\biggr]\,. 
	\nonumber 
\end{equation}
Note that the deformation part contributes to the Hamiltonian as a non-derivative term. 

\medskip 

After adding the Wess-Zumino term to ensure the quantization condition at each site of the spin chain, the total action is given by
\begin{equation}
	\begin{split}
	S &= \dfrac{L}{2}\int\!\! dt d\sigma\, \biggl[- \dfrac{C}{2}(\cosh\rho + \sin\phi\sinh\rho)^{2} + (\cosh\rho - 1)\partial_{t}\phi \\
& \qquad \qquad\qquad \qquad
- \dfrac{\lambda}{16\pi^{2}L^{2}}\bigl[(\partial_{\sigma}\rho)^{2} + \sinh^{2}\rho(\partial_{\sigma}\phi)^{2}\bigr]\biggr]\,,
	\label{e7}
	\end{split}
\end{equation}
through the identification $\xi = 2C$\,.
The action (\ref{e7}) agrees with the Landau-Lifshitz sigma models obtained from the string-theory side, 
up to the S$^1$-contribution. 

\section{Conclusion and Discussion}

We have discussed the correspondence between string theories on warped AdS$_3$ geometries in 
type IIB supergravity and anisotropic deformations of the $sl(2)$ spin chain. 
We have considered three kinds of warped AdS$_3$, 1) space-like warped AdS$_3$,  
2) time-like warped AdS$_3$ and 3) null warped AdS$_3$. 
For the three examples, we have shown the agreement between the fast-moving limit of the relativistic string actions 
and the continuum limit of the corresponding deformed $sl(2)$ spin chains with $sl(2)$ coherent states. 

\medskip 

Let us comment on the contribution of a constant three-form field strength $H_3 = dB_2$\,, 
where $B_2$ is an NS-NS two-form field. The $H_3$ flux is contained in some supergravity solutions including 
the warped AdS$_3\times$S$^1$ geometries.  Then we have to take account of it in our analysis 
because it induces a non-vanishing Wess-Zumino term\footnote{This is different 
from the Wess-Zumino term in the spin-chain side. It exists without the $H_3$ flux.} 
in the string action and, as a result, the Landau-Lifshitz sigma models are also modified. 
So far, we have ignored it for simplicity. We will revisit this issue in another place. 

\medskip 

It is also worth noting on a constant NS-NS $B$-field. 
In the string-theory side, it is basically irrelevant to our analysis because only closed strings are considered. 
Although it seems that a $B$-field, whose two indices are placed on warped AdS$_3$ and S$^1$ respectively, 
gives rise to a non-zero contribution, the S$^1$ part is mostly embedded into a larger space such as S$^3$ and there is no contribution actually. 

\medskip 

In the spin-chain side, the Wess-Zumino term should be added so as to rewrite  
the spin chain Hamiltonian into the Lagrangian form via the Legendre transformation, 
irrespective of $B_2$ and $H_3$ in the string-theory side. 
It can also be interpreted as the Berry phase. This interpretation enables us to see it 
as the contribution of a ``fictitious'' magnetic flux. 


\medskip 

There are various open problems. First, it is of importance to take account of 
higher-order effects in the Landau-Lifshitz sigma models. 
Then it is necessary to consider a long-range generalization of Jordanian deformed spin chains 
like in the BDS spin chain \cite{BDS}. 
The Bethe ansatz is also extended to the all-loop one probably by generalizing the work \cite{KZ}. 
One should check the agreement between the classical energies of the spinning string solutions 
on warped AdS spaces and the energies of the corresponding Bethe string solutions. 

\medskip 

We hope that our work would open a new arena for integrable gauge/string dualities.  

\subsection*{Acknowledgments}

We would like to thank I.~Kawaguchi and T.~Matsumoto for useful discussions. 
The work of KY was supported by the scientific grants from the Ministry of Education, Culture, Sports, Science 
and Technology (MEXT) of Japan (No.\,22740160). This work was also supported in part by the Grant-in-Aid 
for the Global COE Program ``The Next Generation of Physics, Spun 
from Universality and Emergence'' from MEXT, Japan. 

\appendix 

\section{Convention of $sl(2)$ generators}

The abstract Lie algebra $sl(2)$ is formed by the generators $\hat{T}^{a}~(a=0,1,2)$ 
satisfying the following commutation relations,
\begin{equation}
	[\hat{T}^{a}\,,\hat{T}^{b}] = \varepsilon^{ab}_{~~c}\hat{T}^{c}\,. 
	\label{c3}
\end{equation}
Here we have introduced the totally antisymmetric tensor $\varepsilon^{abc}$ with $\varepsilon^{012} = + 1$ and 
the indices are lowered and raised with the $\gamma_{ab} = \textrm{diag}(-1\,, +1\,, +1)$ and its inverse, respectively.   

\medskip 

Then the light-cone notation is defined as
\begin{equation}
	\hat{T}^{\pm} \equiv \dfrac{\hat{T}^{0} \pm \hat{T}^{1}}{\sqrt{2}}\,.
	\label{c2}
\end{equation}
It is useful to list the light-cone components of $\varepsilon^{ab}_{~~c}$ and $\gamma_{ab}$ :
\begin{equation}
	\varepsilon^{-+}_{~~~~2} = +1\,, \quad \gamma_{-+} = \gamma_{+-} = -1\,, \quad \gamma_{22} = +1\,.
	\label{c5}
\end{equation}

\subsection*{For the analysis of string sigma models}

In studying string sigma models on warped AdS$_3$, we use the fundamental representation of the $sl(2)$ generators like 
\begin{equation}
	D(\hat{T}^{0}) \equiv T^{0} = \dfrac{i}{2}\sigma_{2}\,, \quad D(\hat{T}^{1}) \equiv T^{1} = \dfrac{1}{2}\sigma_{1}\,, \quad 
D(\hat{T}^{2}) \equiv T^{2} = \dfrac{1}{2}\sigma_{3}\,,
	\label{c1}
\end{equation}
where $\sigma_{i}\, (i = 1, 2, 3)$ are the standard Pauli matrices. The generators are normalized as 
\begin{equation}
	\textrm{Tr}(T^{a}T^{b}) = \dfrac{1}{2}\gamma^{ab}\,.
\end{equation}

It is helpful to introduce the left-invariant one-form $J$ defined as 
\[
J \equiv g^{-1}dg\,.
\]
It is expanded in terms of the generators $T^{a}$ like
\begin{equation}
	\begin{split}
	J  &= - T^{0}J^{0} + T^{1}J^{1} + T^{2}J^{2} 
\\	&= - T^{+}J^{-} - T^{-}J^{+} + T^{2}J^{2}\,, 
	\label{a8}
	\end{split}
\end{equation}
where $J^{a} \equiv 2\textrm{Tr}(T^{a}J)$\,. 

\subsection*{For the analysis of $sl(2)$ spin chains}

In the analysis of $sl(2)$ spin chains, we use the infinite-dimensional representation of the $sl(2)$ generators 
by following the notation in \cite{SL2}, 
\begin{equation}
	D^{(\infty)}(\hat{T}^{0}) \equiv iS_{0}\,, \quad D^{(\infty)}(\hat{T}^{1}) \equiv - i S_{1}\,, 
\quad D^{(\infty)}(\hat{T}^{2}) \equiv i S_{2}\,.
	\label{c11}
\end{equation}
Here the generators $S_{i}$ are operators acting on a Hilbert space (rather than matrices) 
and satisfy the commutation relations, 
\begin{equation}
	[S_{0}\,, S_{1}] 
				= iS_{2}\,, \qquad
	[S_{1}\,, S_{2}]  
				= - iS_{0}\,,\qquad
	[S_{2}\,, S_{0}] 
				= iS_{1}\,.
	\label{c14}
\end{equation}
We introduce the operators $S_{\pm}$ defined as
\begin{equation}
	S_{\pm} \equiv S_{2}\mp iS_{1 }\,.
	\label{c13}
\end{equation}
Then the commutation relations are rewritten into the following form, 
\begin{equation}
	[S_{-}\,, S_{+}] = 2S_{0}\,, \qquad [S_{0}\,, S_{\pm}] = \pm S_{\pm}\,. 
	\label{c12}
\end{equation}
The light-cone components $\hat{T}^{\pm}$ are represented by 
\begin{equation}
	D^{(\infty)}(\hat{T}^{\pm}) = \dfrac{i}{\sqrt{2}}(S_{0} \mp S_{1})\,.
\end{equation}

\section{The $sl(2)$ coherent states}

In taking the continuum limit of spin chains, one has to use the $sl(2)$ coherent states \cite{Perelomov}. 
We follow the notation in \cite{SL2}.

\medskip 

To define the $sl(2)$ coherent states, let us introduce a unit vector $\vec{n}$ described by
\begin{equation}
	\vec{n} = (n_i) = (\cosh{\rho}, \sinh{\rho}\sin{\phi}, \sinh{\rho}\cos{\phi})\,.
	\label{c15}
\end{equation}
The inner product of $\vec{n}$ is defined as 
\begin{equation}
	\vec{n}^{2} \equiv \eta^{ij}\,n_{i}n_{j} = n_{0}^{2} - n_{1}^{2} - n_{2}^{2} = 1~~~~(\mbox{hyperboloid})\,,
	\label{hyperboloid}
\end{equation}
and the indices are raised and lowered with 
	$\eta^{ij} = \textrm{diag}(+1\,, -1\,, -1)\,.$

\medskip 

By using the unit vector, the $sl(2)$ coherent states are defined as \cite{SL2}
\begin{equation}
	|\vec{n}\rangle \equiv \textrm{e}^{\xi S_{+} - \bar{\xi}S_{-}}|0\rangle\,, \qquad \xi = \dfrac{1}{2}\, \rho\, \textrm{e}^{i\phi}\,,
	\label{c16}
\end{equation}
where $|0\rangle$ is the lowest-weight state with spin 1/2 and satisfies 
\[
S_{0}|0\rangle = \dfrac{1}{2}|0\rangle\,,  \qquad S_{-}|0\rangle = 0\,.
\] 
The expectation value of $S_i$ with the coherent states is evaluated as 
\begin{equation}
	\langle\vec{n}|S_{i}|\vec{n}\rangle = \dfrac{1}{2}n_{i}\,,
	\label{c17}
\end{equation}
and this relation is useful to take the continuum limit of the spin chains. 

\section{The continuum limit of the $sl(2)$ spin chain}

To be self-contained, we shall give a short review on the continuum limit of the $sl(2)$ spin chain   
by sandwiching the spin-chain Hamiltonian in the $sl(2)$ coherent states introduced in Appendix B.

\medskip 

In the abstract form, the Hamiltonian of the $sl(2)$ spin chain with the nearest-neighbor interactions is given by 
\[
H = \dfrac{\lambda}{8\pi^{2}}\sum_{k=1}^LH_{k\,k+1}\,,
\]  
where $L$ is the length of chain and the coefficients are chosen for later convenience\footnote{In the usual AdS$_5$/CFT$_4$\,, 
the prefactor can be computed by perturbative computation in $\mathcal{N}$=4 SYM. However, it is not possible in the present case 
because the field-theory action has not been clarified yet. At most, it can be fixed so as to agree with the string-theory result.}.  
For the detail of the Hamiltonian, see the references \cite{BS,ST,SL2}. This spin chain is often called the XXX$_{-1/2}$ 
Heisenberg spin chain. 

\medskip 

The coherent state for the whole spin chain is given by 
\begin{equation}
	|\vec{n}\rangle \equiv \prod_{k = 1}^{L} |\vec{n}_{k}\rangle \,,
\end{equation}
and the expectation value of $H$ is computed as 
\begin{equation}
	 \langle\vec{n}|H|\vec{n}\rangle =  \dfrac{\lambda}{8\pi^{2}}\sum_{k = 1}^{L} \textrm{log}\biggl(1 - \dfrac{(\vec{n}_{k} - \vec{n}_{k + 1})^{2}}{4}\biggr)\,.
	\label{d3}
\end{equation}
To take a continuum limit, the vector at each site should be replaced as 
\begin{equation}
	\vec{n}_{k} \quad \rightarrow \quad \vec{n}(\sigma) = \vec{n}\left(\dfrac{k}{L}\right)\,, 
\end{equation}
and the difference is expanded in $L\to \infty$ limit like  
\begin{equation}
\vec{n}_{k + 1} - \vec{n}_{k} =  \dfrac{1}{L}\partial_{\sigma}\vec{n} + \mathcal{O}\left(\frac{1}{L^2}\right)\,.
	\label{d4}
\end{equation}
In the limit $L\to \infty$ with $\lambda/L^{2} = $ fixed, the expectation value (\ref{d3}) 
is rewritten as 
\begin{equation}
	\langle\vec{n}| H |\vec{n}\rangle  \quad \longrightarrow \quad 
\dfrac{\lambda}{8\pi^{2}}L\int\! d\sigma\, \dfrac{1}{4L^{2}}\bigl[(\partial_{\sigma}\rho)^{2} 
+ \sinh^{2}\rho(\partial_{\sigma}\phi)^{2}\bigr]\,.
	\label{d5}
\end{equation}

\medskip 

To ensure the quantization condition at each site of the spin chain, it is necessary to add 
the Wess-Zumino term \cite{Kruczenski},  
\begin{equation}
	S_{\rm WZ} = - \dfrac{1}{2} \sum_{k = 1}^{L} \int\! dt\! \int_{0}^{1}\!\! dz \, \vec{n}_{k} \cdot (\partial_{z}\vec{n}_{k} \times \partial_{t}\vec{n}_{k}) = \dfrac{1}{2} \int\!\! dt\, \sum_{k = 1}^{L} (\cosh\rho_{k} - 1)\partial_{t}\phi_{k}\,. 
	\label{d6}
\end{equation}
Thus the continuum limit of the (undeformed) $sl(2)$ spin chain leads to the total action, 
\begin{equation}
	S = \dfrac{L}{2}\int\! dt d\sigma\, \biggl[
(\cosh\rho - 1)\partial_{t}\phi - \dfrac{\lambda}{16\pi^{2}L^{2}}\bigl[(\partial_{\sigma}\rho)^{2} 
+ \sinh^{2}\rho\,(\partial_{\sigma}\phi)^{2}\bigr]\biggr]\,.
	\label{d7}
\end{equation}

\end{document}